\documentclass[aps,prl,twocolumn,showpacs,groupedaddress]{revtex4-1}

\usepackage{graphicx}
\usepackage{amsmath,amssymb}
\usepackage{color,revsymb4-1}

\begin{document}

\title{Quantum Correlations are Tightly Bound by the Exclusivity Principle}

\author{Bin Yan}
\email{yanbin@purdue.edu}
\affiliation{Department of Physics, Purdue University, West Lafayette, Indiana 47906, USA}

\date{11 March 2013}

\begin{abstract}
It is a fundamental problem in physics of what principle limits the correlations as predicted by our current description of nature, based on quantum mechanics. One possible explanation is the ``global exclusivity" principle recently discussed in Phys. Rev. Lett. {\bf110}, 060402 (2013). In this work we show that this principle actually has a much stronger restriction on the probability distribution. We provide a tight constraint inequality imposed by this principle and prove that this principle singles out quantum correlations in scenarios represented by any graph. Our result implies that the exclusivity principle might be one of the fundamental principles of nature. 
\end{abstract}

\maketitle

{\it Introduction.}---Quantum correlations between observables are contextual and nonlocal such that quantum mechanics(QM) is incompatible with either noncontextual hidden variable (NCHV) theories \cite{Specker60, KS67, Bell66} (i.e., predictions of QM cannot be explained by assuming that observables have predefined values which are independent of our choice of measurments) or with local hidden variable (LHV) theories \cite{Bell64} (i.e., as a special case of noncontextuality, results in which are independent of spacelike separated measurements). Noncontextuality (NC) inequalities \cite{CFHR08,KCBS08,Cabello08,SBKTP09,BBCP09} and Bell inequalities \cite{Mermin90,GTHB05,CGR08,CPSS12} (which are a particular type of NC inequalities that require spacelike separated tests) are the basic tools to characterize and reveal quantum correlations. Such correlation inequalities are satisfied by any HCHV and LHV models but are violated by QM. Intriguingly, the maximum quantum violations of these inequalities are bound in a very special way. A fundamental open problem is what is the physical principe that prevents QM from being more contextual \cite{Cabello11, Cabello12} or more nonlocal \cite{PR94,PPKSWZ09,NW09,OW10}?

Recent approaches to the understanding of quantum correlations address this problem into generalized probabilistic theory \cite{Barrett07} and graph theory frameworks \cite{CSW10,FSABCLA12,TAA12,Cabello13}. The main idea is that each correlation inequality can be associated with a group of events, whose relationships can be represented by a graph $G$. $G$ is such a graph that each of its vertices represents an event and two vertices of the graph are adjacent if the corresponding events are mutually exclusive. Correlation inequalities can thus be considered as the sum $S$ of the probabilities of these events. As a convex combination of probabilities, $S$ plays a crucial role in the investigation of quantum correlations since i) it provides generalized forms of correlation inequalities beyond quantum formalism and ii) it might be related directly to the boundaries of quantum probability distributions, as will be discussed in the following.

It has been shown \cite{CSW10,Cabello13} that for a given set of events $\{u_i\}$ with the corresponding graph $G$, the maximum value of $S$ for NCHV and LHV theories is the independent number of $G$, $\alpha(G)$. While the upper bound for $S$ predicted by QM is given by the Lov\'{a}sz number\cite{Lovasz79} of $G$: 
\begin{equation}
\vartheta(G)=Max\sum\limits_{i}|\langle\varphi|v_i\rangle|^2,
\end{equation}
where $|\varphi\rangle$ and $|v_i\rangle$ are unit vectors in Eucledian space and the maximum is taken in any dimensions over all possible $|\varphi\rangle$ and orthogonal representation $\{|v_i\rangle\}$ (which means that each vector $|v_i\rangle$ is assigned to a vertex of $G$ and two vectors are orthogonal if their corresponding vertices are adjacent). $\{|\langle\varphi|v_i\rangle|^2\}$ are QM allowed probability distributions. $S$ always reaches the Lov\'{a}sz number in the minimum dimension that required to produce the orthogonal representation of a graph \cite{supplementary}.
The main idea to understand the origin of the upper bound $\vartheta(G)$ of $S$ is thus to identify ``natural'' information principles, formulated only with the constraints on probability distributions, and also with an intrinsically multipartite structure \cite{MJNANS10,GWAN11}, that prevent stronger correlations than QM.

One possible explanation is from the exclusivity principle(EP) \cite{CSW10,Cabello13}: the sum of probabilities of pairwise exclusive events cannot exceed 1. By pairwise exclusive we mean that if we check any two events of a group of events together, only one of them can occur. On the other hand, a group of events are jointly exclusive means that if we check all the events together, only one of them can be true, no matter how many times and in which order they are tested. Note that Bool's axiom \cite{Boole62} on exclusivity only demands that the sum of the probabilities of jointly exclusive events is less than 1, while in general probability framework pairwise exclusive events are not necessarily jointly exclusive, since tests on different events may affect each other. Consequently, this principle indeed imposes a nontrivial restriction on the potability distributions. This simple principle originally follows from Specker's conjecture on the basic principle of QM(see Ref. \cite{Cabello12_Specker} for a survey), and has been used recently \cite{CSW10,FSABCLA12,TAA12,Cabello13} to investigate its fundamental role in bounding the quantum correlations. 

Denote $\{P_i\}$ as the probabilities for a given set of events $\{u_i\}$. Clearly, according to this principle, the sum of probabilities of any pairwise exclusive events in $\{u_i\}$ cannot exceed 1. Thus, $\{P_i\}$ should at least satisfy
\begin{equation}
\sum\limits_{i\in C}P_i\leq1, \label{constraint1}
\end{equation}
where $C$ is any such subset of $\{u_i\}$ that events in $C$ are pairwise exclusive. Under this constraint, the maximum value of $S$ is the so-called fractional packing number, $\alpha^*(G)$, of the corresponding graph $G$. It has been shown \cite{CSW10} that constraint in inequality (\ref{constraint1}) singles out quantum correlations for a class of scenarios represented by graphs with their fractional packing numbers equal to their Lov\'{a}sz numbers. These scenarios include Bell inequalities for Greenberger-Horne-Zeilinger states \cite{Mermin90} and graph states\cite{GTHB05,CGR08,CPSS12}, and some bipartite Bell inequalities \cite{AGACVMC12,Cabello01} as well as all the state-independent NC inequalities in Ref. \cite{Cabello08}. In a more recent work \cite{Cabello13}, by applying the exclusivity principle to two copies of the Klyachko-Can-Binicioglu-Shumovsky (KCBS) \cite{KCBS08} experiments, Cabello showed that
the upper bound of KCBS inequality for QM is exactly the maximum value allowed by the EP. The quantum violation of KCBS inequality has been experimentally tested with photons \cite{RPCNSMA11,JEAM13}. However, when applying this principle to multiple copies of Clauser-Horne-Shimony-Holt(CHSH) experiments, the answer lead to an open question in graph theory.  It is still unclear whether the maximum quantum violation of CHSH inequality is tightly bound by the EP, or more generally, whether EP can single out quantum correlations in scenarios represented by any graph. Nevertheless, applying only constraint in inequality (2) has been shown \cite{TAA12} to be insufficient to single out quantum probability distributions, even if infinite copies of the original correlation scenarios are taken into account.

In the following we show that, without any additional assumptions, the EP actually has a much stronger restriction on $\{P_i\}$ than inequality (\ref{constraint1}). We provide a tight  constraint inequality on $\{P_i\}$ and show that this principle indeed singles out nature's maximum correlations for any graph.

\begin{figure}
\includegraphics[width=8cm]{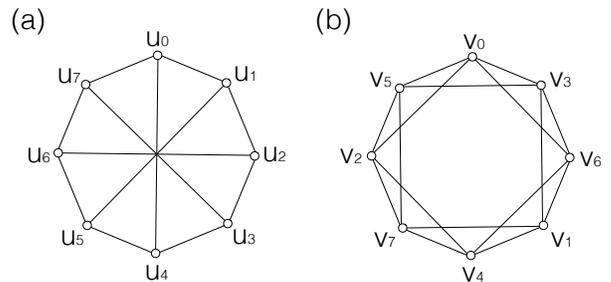}
\caption{\label{fig} (a) Graph \emph{G} of the relationships of the 8 events $\{u_i\}$ involved in CHSH inequality. (b) Graph of the relationships of the 8 events $\{v_i\}$ completely independent with events $\{u_i\}$. Their relationships form the complementary graph of G. This graph is the 8-vertex (1,2)-circulant graph $Ci_8(1,2)$. Joint events $\{u_iv_i\}$ generate the 8-vertex complete graph.}
\end{figure}

{\it Probability distributions allowed by the EP.}---Given a set of events $\{u_i\}$ and the corresponding graph $G$, we now consider another group of events $\{v_i\}$ which are completely independent with $\{u_i\}$. Events $\{v_i\}$ are constructed with the relationships that $v_i$ and $v_j$ are mutually exclusive if $u_i$ is not exclusive with $u_j$. Therefore the corresponding graph of events $\{v_i\}$ is exactly the complementary graph of $G$ [graphs depicted in Fig. (\ref{fig}) are examples of complementary graphs]. We now consider $u_i$ and $v_i$ together as a joint event $u_iv_i$ (which means that event $u_iv_i$ happens if and only if $u_i$ and $v_i$ both happen). Clearly, events $\{u_iv_i\}$ form a pairwise exclusive events set. Donate $\{P_i\}$ and $\{P'_i\}$ the EP allowed probability distributions of $\{u_i\}$ and $\{v_i\}$, respectively. As $u_i$ and $v_i$ are completely independent, the joint probability for event $u_iv_i$ will be $P_iP'_i$.

Note that the joint events $\{u_iv_i\}$ are still real events, despite that they are constructed by jointly viewing two sets of independent events. 
A physical principle should be universal such that the probability distributions of events $\{u_iv_i\}$ must be restricted by the exclusivity principle:
\begin{equation}
\sum\limits_{i} P_i{P'}_i \leq 1. \label{GE_inequality}
\end{equation}
Thus, the constraint on $\{P_i\}$ is that $\{P_i\}$ must satisfy inequality (\ref{GE_inequality}) for \emph{any} EP allowed probability distribution $\{P'_i\}$ on $\overline{G}$. It is a much stronger restriction imposed by the EP since all possible solutions $\{P_i\}$ for this inequality automatically satisfy constraint (\ref{constraint1}). The above inequality is one of the main results in this work. We remark here that the ``global" property assumed in Ref. \cite{Cabello13} actually comes from the universality of a physical principle. Instead of imposing restrictions only on the probability distributions on the given graph $G$ under consideration, as in inequality (\ref{constraint1}), the exclusivity principle exerts constraints upon the whole probability distribution set including all graphs. It demands \emph{compatibility} among different graphs. In other words, EP allowed probability distributions are such that they cannot generate joint probability distributions which do not respect EP. It is in this sense we suggest calling this principle as ``consistent exclusivity". A similar proposal, while under different considerations, has also been discussed in \cite{Henson12,Cabello12_Specker}.

The reason why applying constraint in inequality (\ref{constraint1}) to two copies of KCBS experiments after assuming global exclusivity in Ref. \cite{Cabello13}, singles out that quantum contextuality can be explained as the following: The graph corresponding to KCBS inequality is a pentagon, whose complementary graph is exactly itself. The OR product of two pentagons is a 25-vertex graph which contains five 5-vertex complete graphs. Apply inequality (\ref{constraint1}) to each of the five complete graphs would cover the constraint in inequality (\ref{GE_inequality}). Actually, as a pentagon is self-complementary, inequality (\ref{GE_inequality}) is reduced to a self-constraint inequality $\sum P^2 \leq 1$, which immediately gives us the maximum value of $S$.

In general cases, calculating the upper bound for the sum of $P_i$ comes down to a linear optimization problem subject to constraints in inequality (\ref{GE_inequality}). Now the only problem is that so far we do not know what is exactly the possible probability distribution $\{P'_i\}$ on $\overline{G}$. In fact, EP allowed distributions $\{P'_i\}$ are also bound by inequality (\ref{GE_inequality}) ranging over all EP allowed $\{P_i\}$. This global property, that all probability distributions should be compatible with each other, makes it difficult to estimate the maximum value of $S$ using inequality (\ref{GE_inequality}). Our approach for this problem is to find a subset of all EP allowed $\{P'_i\}$ on $\overline{G}$.

Consider now the QM allowed probability distributions on the given graph $G$ and $\overline{G}$, which (in orthogonal representation) can be written as $\{|\langle\psi|u_i\rangle|^2\}$ and $\{|\langle\varphi|v_i\rangle|^2\}$, respectively. We have the following inequality:
\begin{equation}
\sum\limits_{i} |\langle\psi|u_i\rangle|^2|\langle\varphi|v_i\rangle|^2 \leq 1. \label{QM_inequality}
\end{equation}
This is due to the fact that $\{|u_i\rangle\otimes|v_i\rangle\}$ form an orthogonal system such that $\sum\limits_{i} |\langle\psi|u_i\rangle|^2|\langle\varphi|v_i\rangle|^2 =\sum\limits_{i}|(\langle\psi|\otimes\langle\varphi|)(|u_i\rangle\otimes|v_i\rangle)|^2$ is always less than 1 for arbitrary unit vector $|\psi\rangle \otimes |\varphi\rangle$.

The meaning of inequality (\ref{QM_inequality}) is that all QM allowed probability distributions on graph $G$ and $\overline{G}$ automatically satisfy the restriction imposed by the EP in inequality (\ref{GE_inequality}). Generally, this fact can be seen from a simple property of $\vartheta$ function of any complete graph $G_{complete}$ (in which any two vertices are adjacent):
\begin{equation}
\vartheta(G_{complete}) = 1,
\end{equation}
which means that in QM, the sum of probabilities of any group of pairwise exclusive evens cannot exceed 1. Namely, for any given graph, probability distributions allowed by QM constitute the subset of probability distributions allowed by the EP. 

Note that since the EP might allow probability distributions that can not be realized in the orthogonal representation, it is nontrivial to say that maximum quantum violation is exactly the upper bound imposed by the EP. In other words, we cannot use the orthogonal representation on graph $G$ to estimate the  upper bound of the sum of $P_i$. However, this feature provides a group of possible solutions of $\{P'_i\}$, that is, all QM allowed probability distribution $|\langle \varphi|v_i\rangle|^2$ on $\overline{G}$. $\{P_i\}$ should at least satisfy inequality (\ref{GE_inequality}) when $\{P'_i\}$ adopts these distributions. 

It is worthwhile to mention that  no additional physical assumptions were made in the above discussions about quantum probability distributions.  The point is that probability distributions generated from orthogonal representations provide a possible mathematical solution set for inequality (\ref{GE_inequality}). 

By ranging $\{P'_i\}$ over all unit vector $|\varphi\rangle$ and orthogonal representation $\{|v_i\rangle\}$ on $\overline{G}$, we now get a relatively weaker constraint for $\{P_i\}$: 
\begin{equation}
\sum\limits_{i}P_i|\langle \varphi|v_i\rangle|^2\leq 1. \label{GE_inequality2}
\end{equation}
This constraint actually, as we will see, is sufficient to single out quantum correlations in any graph.

{\it Maximum quantum violation of CHSH inequality.}---We now illustrate how the EP singles out the maximum quantum violation for CHSH inequality \cite{CHSH69}.

The bipartite scenario corresponding to CHSH inequality involves eight events \cite{Cabello13} with their relationships being represented by the graph $G$ depicted in Fig. \ref{fig}(a). 
The upper bounds of CHSH inequality imposed by noncontextual local hidden variable (NCLHV) theory, QM (the bound is known as Tsireson's bound \cite{Cirelson80}) and nonsignaling(NS) \cite{PR94} can be expressed as the following:
\begin{equation}
\sum{P_i}\leq3~^{NCLHV}\leq2+\sqrt{2}~^{QM}\leq4~^{NS}
\end{equation}
where the sum is extended to all the 8 events. Many efforts have been made \cite{PPKSWZ09,NW09,OW10,PW12} to explain why QM stops at Tsireson's bound despite  supraquantum correlations generated from a PR box \cite{PR94} (i.e., a two-party device producing jointly probabilities which satisfy nonsignaling) does not violate nonsignaling. In Ref. \cite{Cabello13}, Cabello showed that the global exclusivity rules out nonsignaling with a lower bound $\frac{8}{\sqrt{5}}\approx3.5778$. (The same value was also obtained in \cite{FSABCLA12} by assuming local orthogonality, which is essentially the same idea as global exclusivity ). 

Here we use the constraint in inequality (\ref{GE_inequality2}) and range all possible orthogonal representations on $Ci_8(1,2)$, the complementary graph of $G$, as depicted in Fig. \ref{fig}(b). Because of the symmetry of $G$, we should expect that the CHSH inequality reaches its maximum value with the eight evens being assigned the same probability P, which is bound by 
\begin{equation}
P\sum|\langle \varphi|v_i\rangle|^2\leq 1. 
\end{equation}
This must be satisfied for any orthogonal representations such that
\begin{equation}
P \leq \frac{1}{\max \sum\limits_{i}|\langle \varphi|v_i\rangle|^2} = \frac{1}{\vartheta(\overline{G})},
\end{equation}
where the maximum ranges over all orthogonal representations on graph $Ci_8(1,2)$ and $\vartheta(\overline{G}) = \max \sum\limits_{i}|\langle \varphi|v_i\rangle|^2$ is the well-known Lov\'{a}sz number for $Ci_8(1,2)$, which is $8-4\sqrt{2}$. We immidiately get the upper bound for CHSH inequality imposed by the EP: $8P_{max}=2+\sqrt{2}$, which is exactly equal to the maximum quantum violation. It is interesting while quite reasonable to see that this number can be given by the complementary graph of the original graph of the 8 events. That is the result of the ``global'' property of exclusivity principle.

{\it Generalization of our result}---We now give a simple proof to show that for correlation inequalities represented by \emph{any} graph, the upper bounds given by the exclusivity principle are exactly the same as QM predicted, such that quantum correlations are tightly bound by the exclusivity principle.

As discussed above, probability distributions $\{P_i\}$ on a given graph $G$ are restricted by inequality (\ref{GE_inequality2}) ranging over all orthonormal representations on $\overline{G}$, the complementary graph of $G$.  For a given normalized vector $|\varphi\rangle$ and orthonormal representation $\{|v_i\rangle\}$, we pick out the minimum value among $\{|\langle \varphi|v_i\rangle|^2\}$, which satisfies
\begin{equation}
(\sum\limits_{i}P_i)\min |\langle \varphi|v_i\rangle|^2\leq\sum\limits_{i}P_i|\langle \varphi|v_i\rangle|^2\leq 1, 
\end{equation}
or
\begin{equation}
\sum\limits_{i}P_i \leq \max \frac{1}{|\langle \varphi|v_i\rangle|^2}. \label{GE_inequaility3}
\end{equation}
Inequality (\ref{GE_inequaility3}) should hold for any normalized vector $|\varphi\rangle$ and orthonormal representation $\{|v_i\rangle\}$ on graph $\overline{G}$ . 
This further gives us:
\begin{equation}
\sum\limits_{i}P_i\leq \min \max\frac{1}{|\langle \varphi|v_i\rangle|^2}, \label{GE_inequality4}
\end{equation}
where the maximum is taken over the given $\{v_i\}$ and $|\varphi\rangle$ and the minimum ranges over all orthonormal representations $\{|v_i\rangle\}$ and unit vectors $|\varphi\rangle$.

The right-hand side of inequality (\ref{GE_inequality4}) equals to the Lov\'{a}sz function for graph $G$ (see $Lemma$ 1 in Ref. \cite{Lovasz79}), which is nothing but the maximum value for QM. 
The equality in (\ref{GE_inequality4}) can hold since the orthogonal representations of graph $G$ generate a possible solution set for $\{P_i\}$ , which can achieve the maximum value $\vartheta(G)$. Interestingly, the tight bound is given by only ranging $\{P'_i\}$ over the subset of all EP allowed distributions on $\overline{G}$. This tells us other possible distributions which are not covered by the orthogonal representations, if there are any, will not give even a lower bound than $\vartheta(G)$.

{\it Conclusion and conjecture.}---
In this work, we have provided further understanding of the global exclusivity principle recently discussed in Ref. \cite{Cabello13}. We show that this principle actually has a much stronger restriction on the probability distribution. Instead of imposing constraint only on the given graph under consideration, this principle imposes restriction on the whole probability structure including \emph{all} graphs. We have provided a constraint inequality and show that this principle indeed singles out quantum correlations represented by any graph. Namely, quantum correlations are tightly bound by this simple principle.

It is not yet unambiguously proven that the probability distributions allowed by QM are all that are allowed by the EP, but the indications are strong. What we can conclude here is that even if the EP may provide probability distribution set which is larger than the quantum set, it will not impose stronger restriction on the quantum correlations revealed by the convex sum of probabilities. It is in this sense we conjecture that the EP not only limits quantum correlations, but also at least almost determines the probability structure of quantum mechanics. Once the probability structure is fully determined, any correlation functions constructed from probabilities, aim at characterizing the quantumness of correlations, are also determined. Further works could focus on such correlation functions, which have other advantages than the correlation inequalities, and generalize them into the graph theory framework.

It is remarkable that our result implies the exclusivity principle, like other physical principles such as uncertainty and nonsignaling, might be one of the fundamental principles of nature.

The author thanks A. Cabello and A. Ac\'{i}n for helpful discussions.


\begin{thebibliography}{33}



\bibitem{Specker60}
 E. P. Specker,
 Dialectica \textbf{14}, 239 (1960).



\bibitem{KS67}
 S. Kochen and E. P. Specker,
 J. Math. Mech. \textbf{17}, 59 (1967).

\bibitem{Bell66}
 J. S. Bell,
 Rev. Mod. Phys. \textbf{38}, 447 (1966).

\bibitem{Bell64}
 J. S. Bell,
 Physics \textbf{1}, 195 (1964).
 

\bibitem{CFHR08}
 A. Cabello, S. Filipp, H. Rauch, and Y. Hasegawa,
 Phys. Rev. Lett. \textbf{100}, 130404 (2008).

\bibitem{KCBS08}
 A. A. Klyachko, M. A. Can, S. Binicio\u{g}lu, and A. S. Shumovsky,
 Phys.~Rev.~Lett. \textbf{101}, 020403 (2008).

\bibitem{Cabello08}
 A. Cabello,
 Phys. Rev. Lett. \textbf{101}, 210401 (2008).

\bibitem{SBKTP09}
 R. W. Spekkens, D. H. Buzacott, A. J. Keehn, B. Toner, and G. J. Pryde,
 Phys. Rev. Lett. \textbf{102}, 010401 (2009).

\bibitem{BBCP09}
 P. Badzi\c{a}g, I. Bengtsson, A. Cabello, and I. Pitowsky,
 Phys. Rev. Lett. \textbf{103}, 050401 (2009).

\bibitem{Mermin90}
 N. D. Mermin,
 Phys. Rev. Lett. \textbf{65}, 1838 (1990).
 
 
 \bibitem{GTHB05}
 O. G\"{u}hne, G. T\'{o}th, P. Hyllus, and H. J. Briegel,
 Phys. Rev. Lett. \textbf{95}, 120405 (2005).

\bibitem{CGR08}
 A. Cabello, O. G\"{u}hne, and D. Rodr\'{\i}guez,
 Phys. Rev. A \textbf{77}, 062106 (2008).

\bibitem{CPSS12}
 A. Cabello, M. G. Parker, G. Scarpa, and S. Severini,
 \eprint{arXiv:1211.4250}.

\bibitem{Cabello11}
 A. Cabello,
 Nature (London) \textbf{474}, 456 (2011).

\bibitem{Cabello12}
 A. Cabello,
 in {\em A Computable Universe},
 edited by H. Zenil
 (World Scientific, Singapore, 2012), Chap.~31.

\bibitem{PR94}
 S. Popescu and D. Rohrlich,
 Found. Phys. \textbf{24}, 379 (1994).


\bibitem{PPKSWZ09}
 M. Paw{\l}owski, T. Paterek, D. Kaszlikowski, V. Scarani, A. Winter, and M. \.{Z}ukowski,
 Nature (London) \textbf{461}, 1101 (2009).

\bibitem{NW09}
 M. Navascu\'es and H. Wunderlich,
 Proc. R. Soc.~A \textbf{466}, 881 (2010).

\bibitem{OW10}
 J. Oppenheim and S. Wehner,
 Science \textbf{330}, 1072 (2010).

\bibitem{Barrett07}
 J. Barrett,
 Phys. Rev.~A \textbf{75}, 032304 (2007)
 
 \bibitem{CSW10}
 A. Cabello, S. Severini, and A. Winter,
 \eprint{arXiv:1010.2163}.

 
 \bibitem{Cabello13}
 A. Cabello,
 Phys. Rev. Lett. \textbf{110}, 060402 (2013)
 
 \bibitem{FSABCLA12}
 T. Fritz, A. B. Sainz, R. Augusiak, J. Bohr Brask, R. Chaves, A. Leverrier, and A. Ac\'{\i}n,
 \eprint{arXiv:1210.3018} 
 
 \bibitem{TAA12}
 T. Fritz, A. Leverrier, and A. B. Sainz,
 \eprint{arXiv:1212.4084}
 
 \bibitem{Lovasz79}
 L. Lov\'{a}sz,
 IEEE~Trans.~Inf.~Theory \textbf{25}, 1 (1979).
 
 \bibitem{supplementary}
As a consequence of this feature, some Bell inequalities cannot reach the maximum value of $S$. For example, Bell inequalities constructed on a pentagon cannot reach its Lov\'{a}sz number $\sqrt{5}$ since the bipartite scenario requires at least a $2\times2$ Hilbert space, while a pentagon reaches $\sqrt{5}$ in three-dimensional Hilbert space. In other words, the enforced bipartite configuration limits our freedom of choosing probability distributions. Such limitation does not come from a physical principle.
 
 \bibitem{MJNANS10}
M. L. Almeida, J-D Bancal, N. Brunner, A. Ac\'{i}n, N. Gisin, and S. Pironio,
Phys. Rev. Lett. \textbf{104}, 230404 (2010)
 
 \bibitem{GWAN11}
 R. Gallego, L. E. W\"urflinger, A. Ac\'{\i}n, and M. Navascu\'es,
 Phys. Rev. Lett. \textbf{107}, 210403 (2011).
 
 
\bibitem{Boole62}
 G. Boole,
 Phil. Trans. R. Soc. Lond. \textbf{152}, 225 (1862).

\bibitem{Cabello12_Specker}
 A. Cabello,
 \eprint{arXiv:1212.1756}
 
 \bibitem{Cabello01}
 A. Cabello,
 Phys. Rev. Lett. \textbf{87}, 010403 (2001).

\bibitem{AGACVMC12}
 L. Aolita, R. Gallego, A. Ac\'{\i}n, A. Chiuri, G. Vallone, P. Mataloni, and A. Cabello,
 Phys. Rev. A \textbf{85}, 032107 (2012).
 
\bibitem{RPCNSMA11}
R. Lapkiewicz, P. Li, C. Schaeff, N. Langford, S. Ramelow, M. Wies\'{n}iak, and A. Zeilinger, 
Nature (London) \textbf{474}, 490 (2011).

\bibitem{JEAM13}
J. Ahrens, E. Amselem, A. Cabello, and M. Bourennane,
\eprint{arXiv: 1301.2887}

 
  \bibitem{Henson12}
J. Henson,
\eprint{arXiv:1210.5978}
 
 \bibitem{CHSH69}
 J. F. Clauser, M. A. Horne, A. Shimony, and R. A. Holt,
 Phys. Rev. Lett. \textbf{23}, 880 (1969).

\bibitem{Cirelson80}
 B. S. Cirel'son [Tsirelson],
 Lett. Math. Phys. \textbf{4}, 93 (1980).
 
 \bibitem{PW12}
 C. Pfister and S. Wehner,
 \eprint{arXiv:1210.0194}.

\end{thebibliography}
\end{document}